\documentclass[12pt]{spieman} 
\usepackage{amsmath,amsfonts,amssymb}
\usepackage{graphicx}
\usepackage{setspace}
\usepackage{tocloft}

\title{Temperature dependence of photo-induced phase segregation in bromide-rich mixed halide perovskites}

\author[a,b,*]{Grigorii Verkhogliadov}
\author[b]{Ross Haroldson}
\author[a]{Dmitry Gets}
\author[a,b]{Anvar A. Zakhidov}
\author[a,c]{Sergey V. Makarov}
\affil[a]{ITMO University, School of Physics and Engineering, Kronverkskiy pr. 49, 197101 St. Petersburg, Russia}
\affil[b]{The University of Texas at Dallas, Physics Department and The NanoTech Institute, Richardson 75080, USA}
\affil[c]{Qingdao Innovation and Development Center, Harbin Engineering University, Qingdao 266000, Shandong, China}

\cftpagenumbersoff{figure}
\cftpagenumbersoff{table} 
\begin{document} 
\maketitle

\begin{abstract}
Mixed halide perovskites undergo phase segregation, manifested as spectral red-shifting of photoluminescence spectra under illumination. In the iodine-bromide mixed perovskites, the origin of the low-energy luminescence is related to iodine-enriched domains formation. Such domains create favorable bands for the induced carrier funneling into them. Despite the phase segregation process is crucial for mixed halide perovskite-based optoelectronics, numerous gaps exist within the understanding of this phenomenon. One such gap pertains to the emergence of temporary and intermediate photoluminescence peaks during the initial stages of phase segregation.However, these peaks appear only within the first few seconds of illumination. Nevertheless, the decreasing temperature may prolong these initial stages. In this work, we carry out a detailed study of the temperature dependence of anion segregation in MAPbBr$_2$I and MAPbBr$_{2.5}$I$_{0.5}$ halide perovskites, to obtain a deeper comprehension of segregation processes, particularly during their initial stages. The temporal evolution of low-temperature photoluminescence reveals the undergoing of the intermediate stage during the segregation process and temperature-related phase transition from orthorhombic to tetragonal phase. To complement the phase segregation study, the temperature dependence of time-resolved photoluminescence spectroscopy is provided, allowing us to estimate the change in the photoluminescence lifetimes for the initial and segregated peaks with temperature. 
\end{abstract}


{\noindent \footnotesize\textbf{*}Grigorii Verkhogliadov,  \linkable{g.verkhogliadov@metalab.ifmo.ru} }

\begin{spacing}{2}   

\section{Introduction}
Since 2009\cite{kojima2009organometal} hybrid perovskites have proven to be an excellent material for optoelectronics, including solar cells (SCs) \cite{Akin2020FAPbI3BasedPS,doi:10.1021/acsenergylett.9b00847,kim2022conformal,mahmoodpoor2022ionic,verkhogliadov2023photoinduced, isikgor2023molecular}, light-emitting diodes (LEDs) \cite{liu2021perovskite,lin2022dual,zhao2023efficient,alahbakhshi2023highly} and various other device types \cite{liang2022high,zhang2021halide, park2022metal}. Additionally, perovskite bandgap can be continuously tuned by incorporating different halides into their structure. By introducing chloride, bromide, and iodide in different proportions to the perovskite composition, the bandgap can be adjusted within the range of 1.5-3.2 eV \cite{li2021bandgap}. Therefore, mixed halide perovskites have gained significant importance in LED applications due to their ability to emit any color of light across the visible range of spectra \cite{zhou2022stabilized,wang2022situ}. Furthermore, mixed halide perovskites are crucial for tandem SCs, where a wide-bandgap perovskite layer is combined with a narrow-band gap absorber. This tandem structure enables the circumvention of the Shockley-Queisser limit imposed on single-junction SCs \cite{chen2022regulating,al2020monolithic}. Particulary, organo-inorganic mixed halide perovskites, such as MAPbBr$_x$I$_{3-x}$, have shown great potential as an efficient wide-bandgap absorber for tandem SCs \cite{tong2022wide,todorov2015monolithic}. Moreover, this perovskite composition has been explored for their application in semi-transparent perovskite SCs \cite{yuan2018semi} and highly responsive photodetectors \cite{mahapatra2021effect}.

 However, mixed halide perovskites exhibit an unwanted phenomenon, which is halide segregation. When exposed to light or an applied bias, the halide anions within the perovskite grains redistribute, resulting in the formation of mono-halide-rich domains with lower bandgap compared to the initial composition. These domains cause a red-shift in the luminescence spectra as photo-generated carriers relax into lower energy states, leading to emission mainly occurring through these low-energy domains. This phenomenon was first observed by Hoke et al. in their pioneer work~\cite{hoke2015}. Moreover, halide segregation is a mainly reversible process, as the perovskite film recovers to the initial homogenized composition if left in the dark \cite{elmelund2019interplay}. Despite being a well-known effect, the exact mechanisms driving halide segregation are not yet fully understood. Three main models have been proposed to explain the driving force behind halide segregation \cite{wang2020phase,knight,choe2021mixed}. The first one is the polaron-based model, which explains halide segregation based on lattice strain induced by the interaction between photogenerated electrons and perovskite ions \cite{bischak2017origin,guo2021photoinduced,guo2020toward,mao2021light,wang2019suppressed}. The second model utilizes thermodynamic potentials to analyze phase diagrams and calculate the preferred state for phase segregation \cite{brivio2016thermodynamic,wang2019suppressed,marchenko2019transferable,lehmann2019phase}. This model provides a combination of total energy calculations and a statistical mechanical approach to analyze the configurational space of the solid-solution composed of mono-halide phases of mixed-halide perovskite \cite{brivio2016thermodynamic}. The third one, known as the bandgap thermodynamic model \cite{kuno2020exactly}. This model describes the kinetic process leading to the redistribution of ions by photoexcited charge carriers, which accumulate and recombine in lower bandgap domains  \cite{draguta2017rationalizing,brennan2017light}.  Kinetic Monte Carlo simulations have been employed to improve the bandgap thermodynamic model, aiming to achieve improved consistency with experimental observations through the direct connection of the individual anions' migration with the evolution of photoluminescence (PL) and absorption spectra \cite{ruth2018vacancy}. However, recent work \cite{suchan2023multi} has revealed that the previously described thermodynamic models only rationalize excitation equal to 1 sun in the initial stage of phase segregation. The final equilibrium state achieved during laser irradiation reveals that the energy of the average charge-carrier density does not adequately explain the extent of the segregation observed. 

Therefore, phase segregation in mixed-halide perovskites is a complex phenomenon that is strongly influenced by various factors, such as the intensity of illumination, the amount of defects, and the composition of the perovskite. Although halide segregation is usually represented as a straightforward red-shift of a PL spectrum toward lower energies, this process may also involve the formation of intermediate peaks between the initial (host) perovskite PL peak and the fully segregated one \cite{mao2019visualizing}. Additionally, the bromide-rich composition of organo-inorganic halide perovskite demonstrates the emergence of temporary low-energy PL associated with the formation of nano-domains within the first few seconds of illumination \cite{suchan2020complex}. Consequently, a comprehensive understanding of the initial processes of phase segregation is crucial for elucidating its underlying mechanism. In this regard, low-temperature measurements provide a valuable opportunity to gain insight into the phase segregation process. Such measurements facilitate the observation of all migration processes at lower rate~\cite{lehmann2019phase,liu2018temperature,lee2017temperature,gautam2020reversible,pavlovetc2021distinguishing}, which is essential for observation of the initial stages of the photo-segregation.

In this study, we investigate the temperature dependence of halide segregation in bromide-rich MAPbBr$_x$I$_{3-x}$ perovskite for \textit{x}=2.5 and 2, which are one of the most popular mixed perovskites. Previous studies have not offered a comprehensive investigation into the dynamics of temporary and intermediate PL formation on a timescale exceeding a few seconds in duration. Here, we aim to address this gap by analyzing the evolution of spectra over time under laser irradiation at various temperatures. This approach not only enables us to uncover the phase transitions occurring within the mixed halide perovskite structure, but also facilitates the observation of temporary and intermediate PL peaks during the initial stages of photo-induced segregation for an extended timescale. The decreasing of the temperature allows for achieving a lower rate of ionic migration and prolongs the first stages of segregation from seconds to minutes. In addition to the spectral measurements, we also perform temperature-dependent time-resolved photoluminescence (TRPL) measurements, which enable the evaluation of the radiative and non-radiative lifetime of mixed halide perovskite at different temperatures for both the initial and segregated peaks. These can provide valuable insights into the carrier recombination rate within iodine-enriched domains of the mixed-phase perovskites.

\section{Results and discussion}

\begin{figure}
    \centering
    \includegraphics[width=1\textwidth]{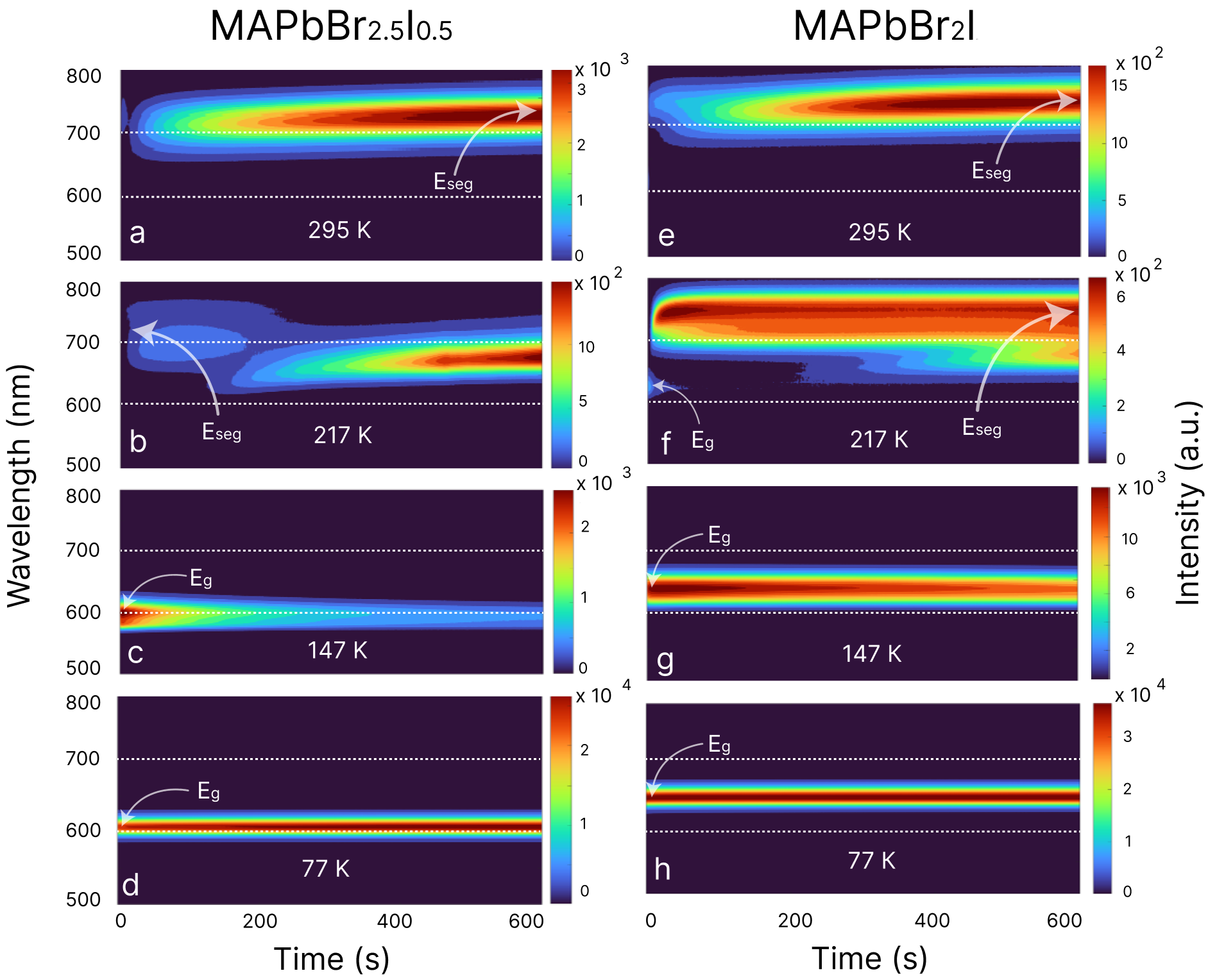}
    \caption{2D maps of PL spectra at different temperatures, the color represents the luminescence intensity, intensity of the laser 100 mW/cm$^2$, integration time vary from 50 ms to 200 ms depends on temperature. a) MAPbBr$_{2.5}$I$_{0.5}$ perovskite at 295 K b) MAPbBr$_{2.5}$I$_{0.5}$ perovskite at 217 K c) MAPbBr$_{2.5}$I$_{0.5}$ perovskite at 147K d) MAPbBr$_{2.5}$I$_{0.5}$ perovskite at 77 K. e) MAPbBr$_{2}$I perovskite at 295 K f) MAPbBr$_{2}$I perovskite at 217 K g) MAPbBr$_{2}$I perovskite at 147K h) MAPbBr$_{2}$I perovskite at 77 K.}
    \label{fig1}
\end{figure}

This study focuses on mixed halide perovskites containing methylammonium as an organic cation and a mixture of iodide and bromide in ratios of 1:5 and 1:2 (MAPbBr$_{2.5}$I$_{0.5}$ and MAPbBr$_2$I respectively). The bromide-rich compositions exhibit intriguing segregation behavior, as depicted in Figure \ref{fig1}. Halide segregation refers to the appearance of iodine-enrich regions in a film, as previously mentioned. However, in bromide-rich compositions, this process is not a simple linear shift of the spectra from the initial peak to the segregated one. For instance, in Figure \ref{fig1}a for MAPbBr$_{2.5}$I$_{0.5}$ perovskite composition at room temperature, after the emergence of the segregated peak at approximately  720 nm, another peak arises around 700 nm between the initial and segregated peaks. These PL spectra gradually red-shifts to the fully segregated position over time. Similar PL spectra change over time in bromide-rich compositions were previously demonstrated in works made by Suchan et al\cite{suchan2020complex}. The redistribution of halide in perovskite domains and related PL spectra changes during their formation can be divided into three stages. The first one is an initial stage, before any light illumination (0 seconds in Figure \ref{fig1}a), followed by the intermediate stage, which occurs at the beginning of illumination (from 0 to approximately 20 seconds in Figure \ref{fig1}a), and third stage, which starts from tenth seconds until the final halide redistribution in perovskite film in Figure \ref{fig1}a. During the third stage, spectra changing can be described as a linear red-shifting of peak position. However, in the intermediate stage, spectra changes are more complex. During this stage, the red-shift of PL spectra occurs initially, followed by a subsequent blue-shift of the spectra (PL spectra obtained at various time points during the intermediate stage are provided in supplementary notes).

In order to gain a more precise understanding of the intermediate-stage processes, we conducted low-temperature measurements of halide segregation. Figure \ref{fig1}b shows the same composition at 217K, where the emergence of an intermediate peak between the initial and segregated peaks is more readily observable over time. As the temperature decreases to 147K and 77K (Figures \ref{fig1}c and \ref{fig1}d), PL segregation is suppressed at a 1 sun laser intensity and irradiation time of 10 minutes.

Similar to the previous composition, the appearance of intermediate peaks can be clearly observed in Figures \ref{fig1}e and \ref{fig1}f. Additionally, Figure \ref{fig1}f illustrates that, during the initial few seconds of illumination, only the initial peak of the spectra is present, but the segregated peak around 750 nm appears almost immediately. Subsequently, after approximately 2 minutes of illumination, an intermediate peak around 700 nm arises, followed by a second intermediate peak around 650 nm after roughly 5 minutes of illumination. These peaks also exhibit redshifts with time under continuous laser irradiation. Consistent with the previous composition, no segregation is detected at lower temperatures of 147K and 77K, as demonstrated in Figures \ref{fig1}g and \ref{fig1}h.

\begin{figure}
    \centering
    \includegraphics[width=1\textwidth]{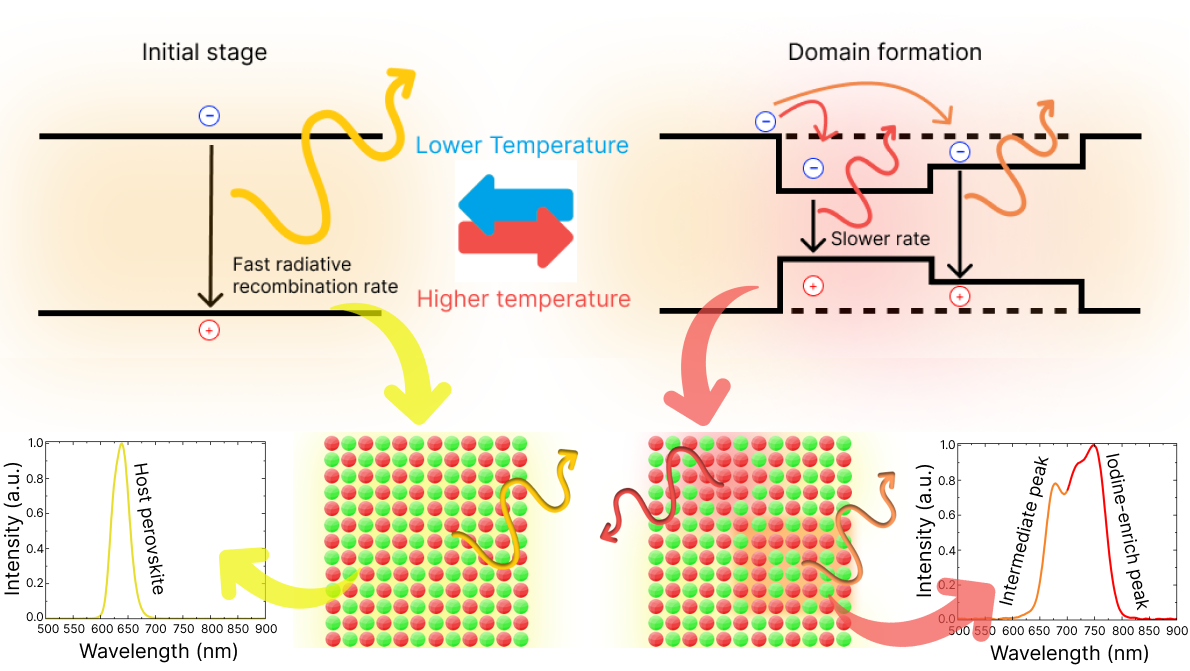}
    \caption{Domain formation on band diagram and on crystal structure schematic with temperature increasing with demonstrating PL spectra in both initial stage and in intermediate stage with formed domains with different halide composition.}
    \label{fig2}
\end{figure}

During the intermediate stage, the emergence of intermediate peaks is related to the formation of minuscule domains. \cite{suchan2023multi}. The sequence of intermediate peaks appearance may be linked to the migration rate of halide ions. Notably, iodine exhibits a much higher migration rate than bromide \cite{mcgovern2020understanding,futscher2019quantification,lee2023situ}, enabling it to reach perovskite vacancies and create iodine-enrich areas. However, bromide ions are also capable of migration and can eventually occupy these vacancies. As our film contains an excess of bromide relative to iodine, some bromide ions can take up these vacancies over time and form areas with distinct bromide-to-iodide ratios compared to the initial composition. At low temperatures, while the formation of domains was ceased due to the suppression of ion migration in the perovskite, the band diagram and crystal structure remain in their initial state. However, an increase of temperature stimulates halide redistribution, leading to changes in band structure, as it is shown in Figure~\ref{fig2}. Formation of domains with lower bandgap, which act like trap states for carriers, leading to the appearance of red-shifted peaks on spectra (PL spectra are also shown in Figure~\ref{fig2}).

\begin{figure}
    \centering
    \includegraphics[width=1\textwidth]{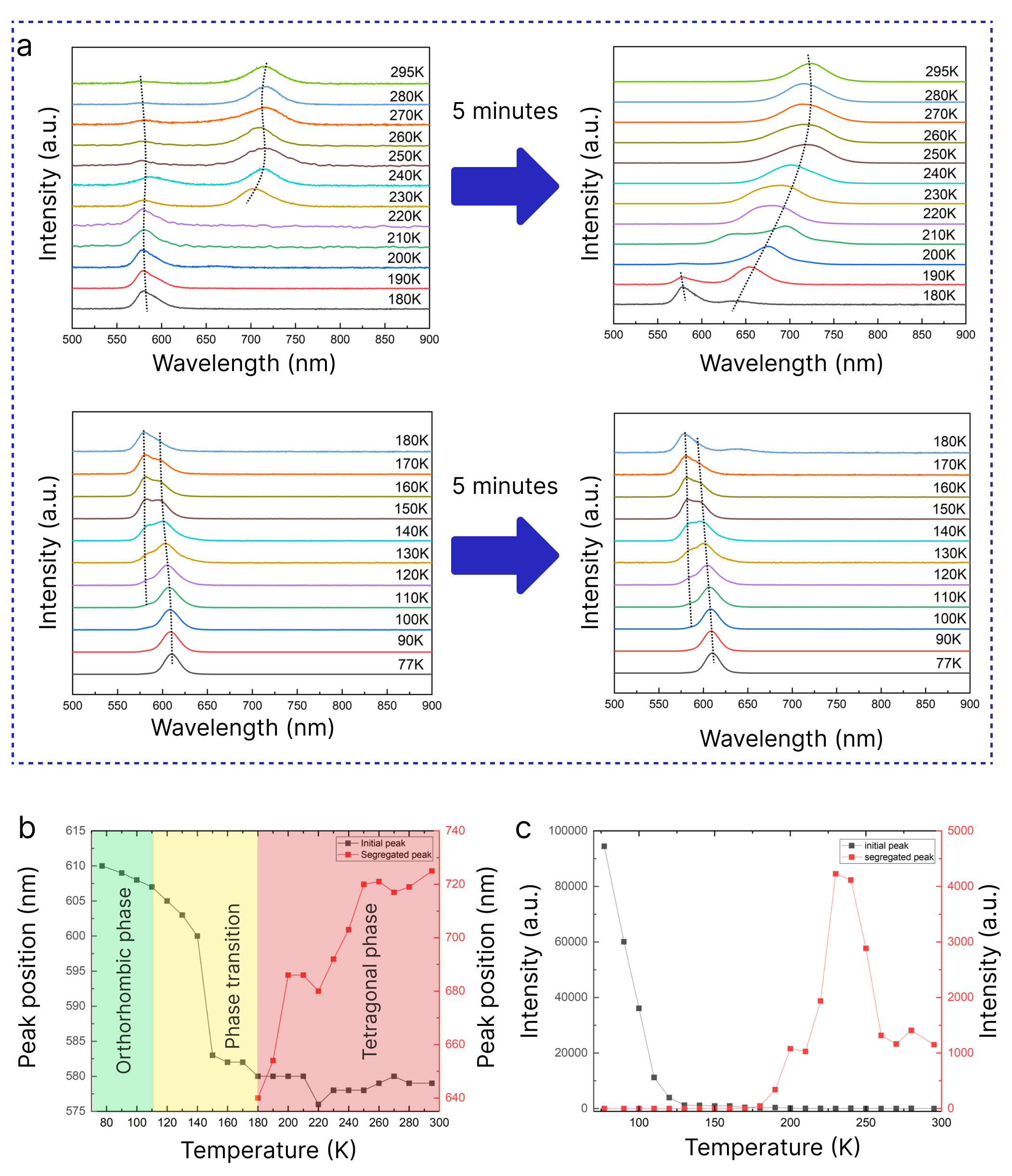}
    \caption{MAPbBr$_{2.5}$I$_{0.5}$ perovskite composition. a) PL spectra in the beginning and after 5 minutes of laser illumination with 200 mW/cm$^2$ intensity for temperature range from 77 K to 295K with 10 K step. b) Initial and segregated peak position for the temperature range from 77 K to 295 K with 10 K step. Green area represents perovskite in the orthorhombic phase, yellow area -- the transition from orthorhombic to tetragonal phases, red area -- the tetragonal phase. c) Initial and segregated peak intensity for the temperature range from 77 K to 295 K with 10 K step.}
    \label{fig3}
\end{figure}

Furthermore, it is evident that the peak position shifts for both compositions as the temperature increases from 77 K to 147 K.  To investigate the temperature dependence of PL, spectra were measured for 5 minutes under laser exposure at different temperatures with 10 K step. Figure \ref{fig3}a shows these spectra at the beginning of illumination and after 5 minutes for MAPbBr$_{2.5}$I$_{0.5}$. At 110 K another peak near the initial one begins to grow and at approximately 180 K, it becomes dominant, indicating a phase transition in the perovskite structure. Below 110 K, the perovskite is in the orthorhombic phase, but at 110 K, a gradual transition to the tetragonal phase commences \cite{liu2018temperature}. At 180 K, the perovskite is mostly in the tetragonal phase. Therefore, in the temperature range between 110 K and 180 K, two phases of perovskite are present in the composition. Additionally, 180 K is a threshold temperature for phase segregation, as the red-shifted peak becomes noticeable after 5 minutes of irradiation with an intensity of 200 mW/cm$^2$. Although halide segregation in this composition only starts after the completed phase transition from orthorhombic to tetragonal, segregation in the orthorhombic phase is possible \cite{li2019temperature,hu2021linking}. This threshold temperature is likely related to increase in ion mobility with temperature growth.

Figure \ref{fig3}b illustrates the variation in peak position with temperature for both the initial and segregated peaks. A shift of 25-30~nm in the peak position between 110 K and 180 K indicates a phase transition, after which the initial peak position stabilizes. In contrast, the segregated peak shifts towards the infrared region with increasing temperature until approximately 240 K. This can be attributed to an increase in the diffusion coefficient and ions' migration rate \cite{lai2018intrinsic,pols2021atomistic}. At lower temperature, ions require more time to form iodine-enrich domains. As temperature increases, all migration processes occur at a faster rate, leading to a notable red-shift in the peak position in the same period of time. Figure \ref{fig3}c demonstrates the changes in PL intensity with temperature. The initial peak intensity decreases with temperature growth, whereas the segregated peak intensity increases from 180 K to 240 K and then begins to decrease. The temperature at which the segregated peak reaches its maximum intensity corresponds to the red-shifting limit of the perovskite composition. Within the temperature range of 180 K to 240 K, perovskite resides in an intermediate stage of segregation, wherein both peak position and intensity experience a significant and rapid alteration. This growth in photoluminescence intensity during the intermediate stage of segregation is attributed to the formation of minuscule domains. These nano-domains are enveloped by a higher bandgap host perovskite matrix, thereby impeding charge carriers from leaving favorable bands and preventing non-radiative recombination. It can be inferred that 5 minutes of laser irradiation at 200 mW/cm$^2$ at 240 K is sufficient to pass the intermediate stage of segregation.  As a result, the decrease in intensity with temperature becomes more pronounced after this temperature point. However, intensity vs temperature dependency of the segregated peak from the area at the third stage of segregation should exhibit the same trend as the dependency observed for the initial peak. The elevation in temperature has the potential to activate more channels for non-radiative recombination, which is expressed in a decrease in the PL intensity.

\begin{figure}
     \centering
    \includegraphics[width=1\textwidth]{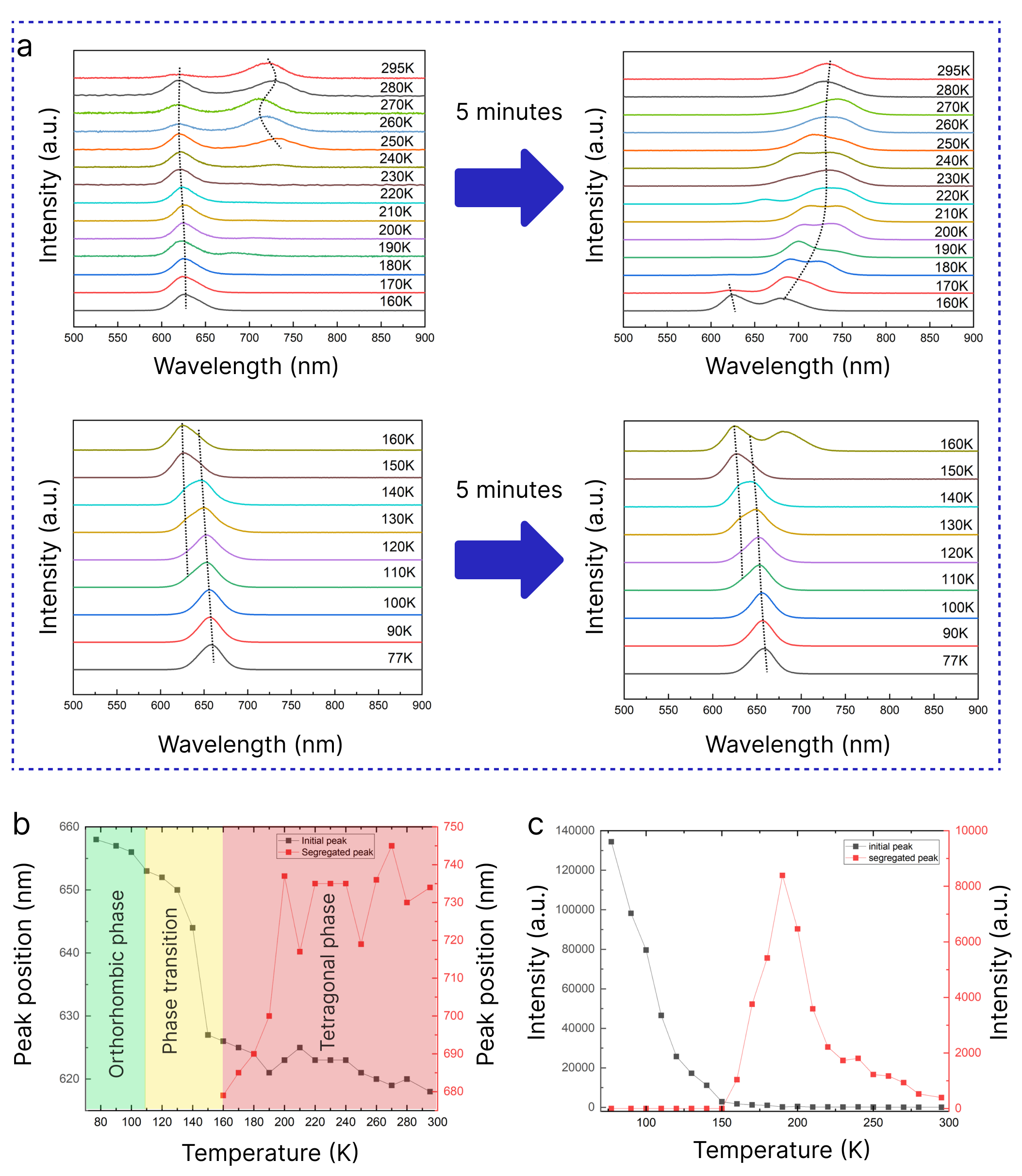}
    \caption{MAPbBr$_2$I perovskite composition. a) PL spectra in the beginning and after 5 minutes of laser illumination with 200 mW/cm$^2$ intensity for temperature range from 77 K to 295K with 10 K step. b) Initial and segregated peak position for the temperature range from 77 K to 295 K with 10 K step. Green area represents perovskite in the orthorhombic phase, yellow area -- the transition from orthorhombic to tetragonal phases, red area -- the tetragonal phase. c) Initial and segregated peak intensity for the temperature range from 77 K to 295 K with 10 K step.}
    \label{fig4}
\end{figure}

Figure \ref{fig4}a shows the PL spectra changing of MAPbBr$_2$I, the second bromide-rich composition. A peak emerges at 110 K, corresponding to a distinct phase of perovskite as well. This peak becomes dominant at 150-160 K, similar to the behavior observed in the MAPbBr$_{2.5}$I$_{0.5}$ composition. The segregation threshold temperature for this composition is 160 K at a laser intensity of 200 mW/cm$^2$. Moreover, in Figures \ref{fig4}b and \ref{fig4}c we observe a similar tendency to the previous composition. The initial peak shifts between 110 K and 160 K, and then stabilizes, while the segregated peak red shifts from 160 K to 200 K and stabilizes at approximately 740 nm. It is noteworthy that both the polaron model and the thermodynamic potential model suggest that the position of the fully segregated peak shifts towards longer wavelengths as the temperature increases \cite{bischak2017origin}. However, in this study, we observed a temperature-independent fully segregated peak position, which is in good agreement with previously published findings \cite{pavlovetc2021distinguishing,kuno2020exactly}. Remarkable, that MAPbBr$_2$I segregated peak is more red-shifted compare to MAPbBr$_{2.5}$I$_{0.5}$, which may be attributed to the presence of iodine ions in the mixture that can form iodine-enrich areas. For MAPbBr$_{1.5}$I$_{1.5}$ segregated peak reaches 750-755 nm, which is shown in supplement notes. The intensity vs temperature plots exhibit a similar decrease in the initial peak with increasing temperature, and the segregated peak also shows a correlation with temperature change comparable to that observed in the MAPbBr$_{2.5}$I$_{0.5}$ composition.

As mentioned earlier, the polaron and thermodynamic potential models face difficulties in explaining the temperature independence of the position of the fully segregated peak. However, the bandgap thermodynamic model provides a more comprehensive description of the phase segregation mechanism. It successfully considers factors such as the temperature independence of the fully segregated peak's position and the intensity threshold for halide segregation. Nevertheless, this model does not perfectly fit all experimental data \cite{suchan2023multi}. Previous studies have mainly focused on investigating halide segregation for a 1:1 bromide-iodine ratio or compositions with an excess of iodine \cite{ruth2018vacancy,brennan2017light,draguta2017rationalizing}. However, the emergence of the temporal and intermediate peaks that we are studying here occurs in bromide-rich compositions. Furthermore, these peaks only appear for a very short period at the beginning of illumination. The precise modeling of intermediate peak formation may hold the key to understanding the mechanisms underlying phase segregation in mixed-halide perovskites.

In order to illustrate the PL decay process, TRPL measurements were  conducted on two bro-mide-rich compositions, namely MAPbBr$_{2.5}$I$_{0.5}$ and MAPbBr$_{2}$I. The TRPL curves were measured within a temperature range of 77 K to 300 K with a 20 K increment. Bi-exponential fitting was applied to extrapolate radiative lifetime (t$_{rad}$) and non-radiative lifetime (t$_{srh}$) from Pl decay curves at each temperature \cite{ding2018role,chirvony2020interpretation}. Further details regarding the fitting formula can be found in supplementary notes. Figure \ref{fig5}a presents the PL decays for MAPbBr$_{2}$I perovskite composition for three different temperatures, namely 77 K, 140 K and 200 K. It can be observed that t$_{rad}$ increases from 1.97 ns at 77 K to 2.36 ns at 140 K, while t$_{srh}$ remains approximately constant (5.28 ns at 77 K vs 5.44 ns at 140 K). This leads to an increase in PL quantum yield within this temperature range. However, at 200 K, t$_{rad}$ remains relatively constant at 2.32 ns, while t$_{srh}$ increases to 6.83 ns. The inset in Figure \ref{fig5}a  reveals the emergence of a segregated peak at 200 K in the PL spectra.

To distinguish the impact of the initial peak and segregated peak on the lifetimes separately, 700 nm short-pass and long-pass filters were employed to divide the initial peak and segregated peak. Figure \ref{fig5}b displays PL decay from the initial peak only. The main contribution to the PL decay at 200 K is attributed to the initial peak, so t$_{rad}$ and t$_{srh}$ remain almost similar to full spectra lifetimes (2.11 and 6.57 ns, respectively) and remain relatively constant with increasing temperature up to 240 K (2.05 ns and 6.76 ns, respectively). However, with further increase of the temperature both t$_{rad}$ and t$_{srh}$ decrease dramatically, reaching 0.82 ns and 3.25 ns, respectively, at room temperature. This can be explained by the suppression of the initial peak at room temperature due to halide segregation (spectra are shown in the inset to Figure \ref{fig5}b).

In Figure \ref{fig5}c, it can be observed that the segregated peak exhibits a higher t$_{rad}$ of 3.58 ns at 200 K compared to the full spectra t$_{rad}$, but it also displays a higher t$_{srh}$ of 7.44 ns. As the temperature increases to 220 K, t$_{rad}$ rises to 3.92 ns, while t$_{srh}$ experiences a more substantial increase to 11.05 ns.  Furthermore, the segregated peak becomes more pronounced at 220 K, as shown in the inset of Figure \ref{fig5}c.  At 260 K t$_{rad}$ decreases to 2.95 ns, but t$_{srh}$ remains relatively constant at 10.58 ns. However, at room temperature, both t$_{rad}$ and t$_{srh}$ significantly decrease to 1.51 ns and 5.05 ns, respectively. The initial increase in t$_{rad}$ of the segregated peak is associated with the process of domain formation during the intermediate stage of segregation, which leads to an increase in the excited state and hence the intensity of PL (Figure \ref{fig4}c). However, after this point, the temperature effect becomes dominant, resulting in a decrease in t$_{rad}$. 

\begin{figure}
    \centering
    \includegraphics[width=0.95\textwidth]{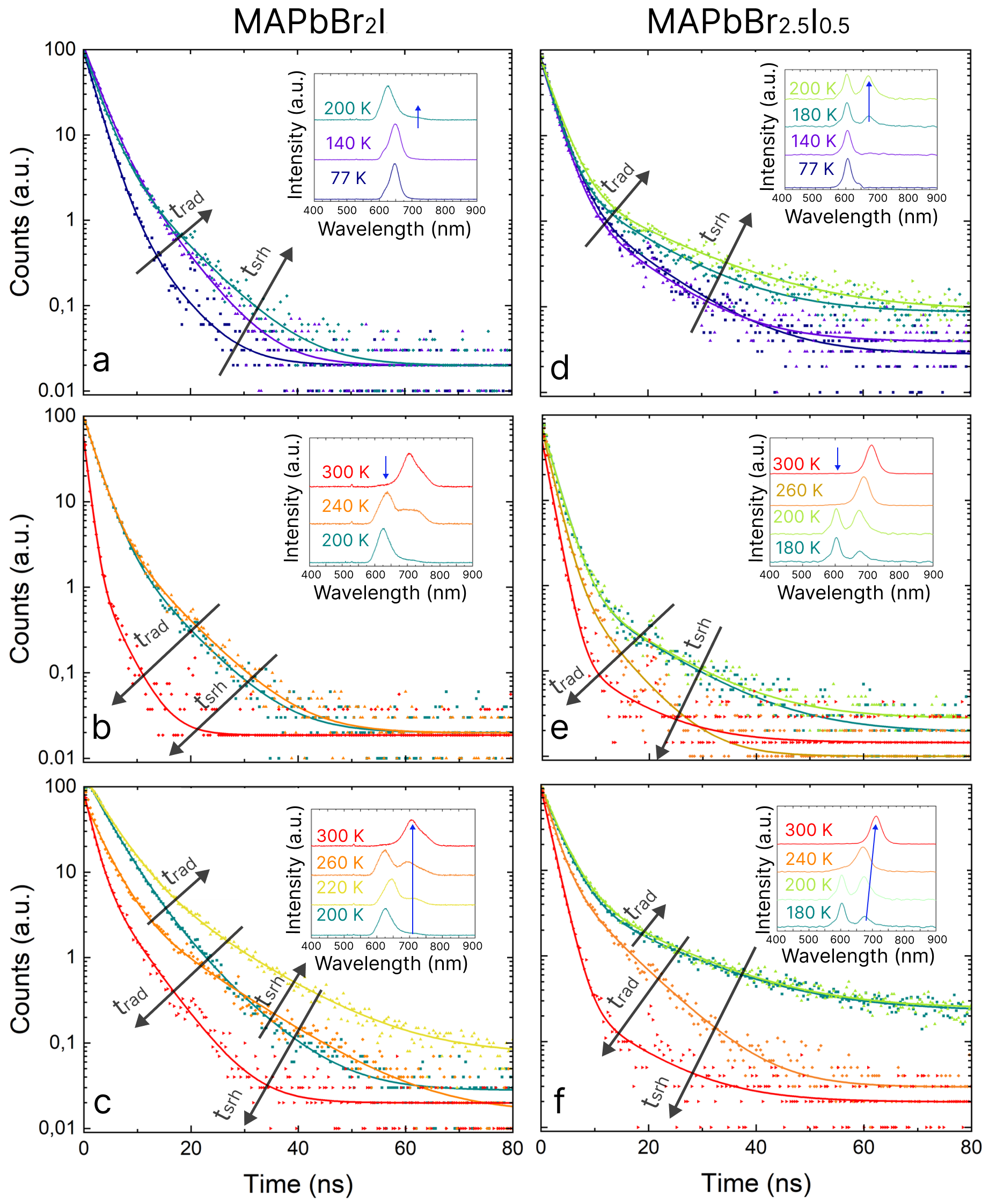}
    \caption{PL decays experiment (dots) vs fitting (solid line) at different temperatures (represented by color), inserts demonstrates PL spectra at the same temperature. a) PL decays for MAPbBr$_2$I at different temperatures from whole spectra. b) PL decays for an initial peak at different temperatures for MAPbBr$_2$I perovskite. c) PL decays for segregated peak at different temperatures for MAPbBr$_2$I perovskite. d)  PL decays for MAPbBr$_{2.5}$I$_{0.5}$ at different temperatures from whole spectra. e) PL decays for initial peak at different temperatures for MAPbBr$_{2.5}$I$_{0.5}$ perovskite. f) PL decays for segregated peak at different temperatures for  MAPbBr$_{2.5}$I$_{0.5}$ perovskite.}
    \label{fig5}
\end{figure}

MAPbBr$_{2.5}$I$_{0.5}$ composition displays a similar behavior. The PL decays for this composition are presented in Figures \ref{fig5}d, e and f, where Figure \ref{fig5}d demonstrates TRPL curves from the entire spectra, and Figure \ref{fig5}e and f display PL decays with contribution from the initial peak and segregated peak, respectively, divided by 650 nm short pass and long pass filters. the lifetimes for each temperature can be found in tables S1 and S2 in supplementary notes. 

The variation of PL lifetime can be divided into several stages. Considering the case of MAPbBr$_2$I composition, the first stage occurs within a temperature range of 77 K to 140 K, during which the lifetime increases, as it was described below (for t$_{rad}$ from 1.97 to 2.36 ns). It is worth noting that 140 K marks the approximate temperature at which the phase transition for the MAPbBr$_2$I composition begins. The second stage corresponds to a further increase in temperature, up to 180-200 K, which is the threshold temperature for halide segregation. During this stage, the lifetime decreases (for t$_{rad}$ from 2.36 to 2.01 ns), providing evidence for the correlation between the perovskite phase and recombination rates \cite{liu2018temperature,yi2020correlation}. However, once halide segregation starts to occur (above 180 K), the lifetime begins to rise again until approximately 240 K (for t$_{rad}$ from 2.01 to 2.69 ns). This increase in lifetime can be attributed to the formation of minuscule iodine-enriched domains, where carriers require more time to localize and recombine. The impact related to the formation of these domains dominates over the lifetime decrease observed in the tetragonal phase with increasing temperature during the previous stage. In the final stage, from 240 K until room temperature, the lifetime decreases once again (for t$_{rad}$ from 2.69 to 1.94 ns). Interestingly, in mono-halide inorganic perovskites, the orthorhombic phase also exhibits a rise in the lifetime with increasing temperature \cite{yi2020correlation}. However, the transition from the orthorhombic phase to the monoclinic phase results in a reverse trend in the temperature-dependent lifetime. Therefore, in our case, although the phase transition affects the lifetime, the influence of halide segregation is also present. Consequently, this leads to the emergence of four distinct intervals wherein the PL lifetime experiences alternating increases or decreases as the temperature rises.

\section{Conclusion}
In this study, we have presented an experimental analysis of phase segregation processes in bro-mide-rich mixed halide perovskite at the broad range of temperatures from 77~K to 295~K. Our spectral measurements provide valuable information on temperature-related ranges of phase transition in mixed halide perovskites. By reducing the temperature, we have prolonged the intermediate stage of halide segregation from seconds at room temperature to minutes at 217~K. The deceleration of the domain formation process with decreasing temperature enables us to identify the emergence of intermediate peaks that were not properly observable at room temperature before.

The emergence of the aforementioned temporary peaks in bromide-rich perovskites is associated with the formation of minuscule domains at the beginning of the phase segregation process. Nevertheless, an excess of bromide within the perovskite film results in the formation of nano-domains possessing a distinct halide ratio when compared to the host perovskite composition or a fully segregated one. The sequence of domain emergence is closely related to the migration rate of various halides, with iodine domains predominantly forming first, followed by domains where an increasing amount of bromide is incorporated. Additionally, temperature-dependent TRPL measurements have complemented the obtained results, providing two characteristic time-scales of radiative t$_{rad}$ and defect-assisted t$_{srh}$ recombinations which are also functions of temperature. We divided the alternating increase and decrease of PL lifetime into four distinct intervals that are linked to the temperature at which the phase transition occurs in mixed halide perovskite, as well as the threshold temperature for phase segregation. 

As a result, the acquisition of experimental data pertaining to halide segregation during its intermediate stage, in conjunction with TRPL measurements, can provide significant insights into the underlying processes driving this phenomenon. The existing models, such us polaron and thermodynamic potential models, encounter difficulties in accurately describing certain temperature-dependent experimental data. However, the bandgap thermodynamic model exhibits improved conformity with experimental data, albeit not entirely fitting in instances where illumination intensity equals 1 sun. Therefore, this research could aid in the development of accurate models, facilitating a deeper comprehension and identification of strategies to suppress phase segregation. Our findings pertaining to the dynamics of the intermediate stage of phase segregation could help to develop a complete phase segregation theory, which has the potential to deliver a basis for the production of stable mixed halide SCs and LEDs.

\section{Methods}
MAPb$_x$I$_{3-x}$ samples were synthesized using stoichiometric mixtures of MAI (99.99\% from Greatcell solar materials), MABr (99.99\% from Greatcell solar materials), PbBr$_2$ (99.999\%, metal basis, from Alfa Aesar) in a mixture dimethylformamide and dimethyl sulfoxide in ratio 7:3. All reactions were performed in a glovebox under nitrogen atmosphere. 

Thin films were fabricated by the spin-coating method using a two-step spin-cycle with 1000 rpm for the first step and 3000 rpm for the second one with toluene as an antisolvent to obtain films with thicknesses of 300-350 nm. Film thickness  was determined using a profilometer Ambios technology XP-1. 

All PL measurements were performed in a probe station (Desert Cryogenics) under vacuum conditions. A 405 nm continuous wave laser from Thorlabs was used as the illumination source. For low-temperature measurements,  liquid nitrogen was utilized to achieve the required low temperatures. An Ocean Optics spectrometer was employed for all spectral measurements. Optical pumping for TRPL measurements was achieved using a  Pharos, Light conversion pulse laser with a wavelength of 532 nm, pulse duration of 220 fs, frequency of 250 kHz, and an intensity of laser radiation of 35 mW/cm$^2$.

\subsection* {Acknowledgments}
This work was supported by the Ministry of Science and Higher Education of the Russian Federation (Agreement No. 075-15-2021-589) and partially by the Welch Foundation of Texas (grant AT-1617).

\subsection* {Supporting Information}

Supporting information contains 2D maps of PL spectra for MAPbBr$_{x}$I$_{3-x}$, (x=1.5; 1; 0.5); t$_{rad}$ and t$_{srh}$ at different temperatures; spectral changing during intermediate stage; thin film characterization with atomic force microscopy, absorption spectroscopy and PL quantum yield measurements; information about experimental setups.

\bibliography{article}  
\bibliographystyle{spiejour}   

\newpage
\section{Supplementary notes}

Figures \ref{figs1} demonstrate 2D photoluminescence (PL) maps for MAPbBr$_{1.5}$I$_{1.5}$ perovskite composition at 4 different temperatures. This composition do not demonstrate formation of intermediate stages during segregation processes like bromide-rich composition, which is in good agreement with previously published data \cite{suchan2020complex,suchan2023multi}.

\begin{figure}
    \centering
    \includegraphics[width=1\textwidth]{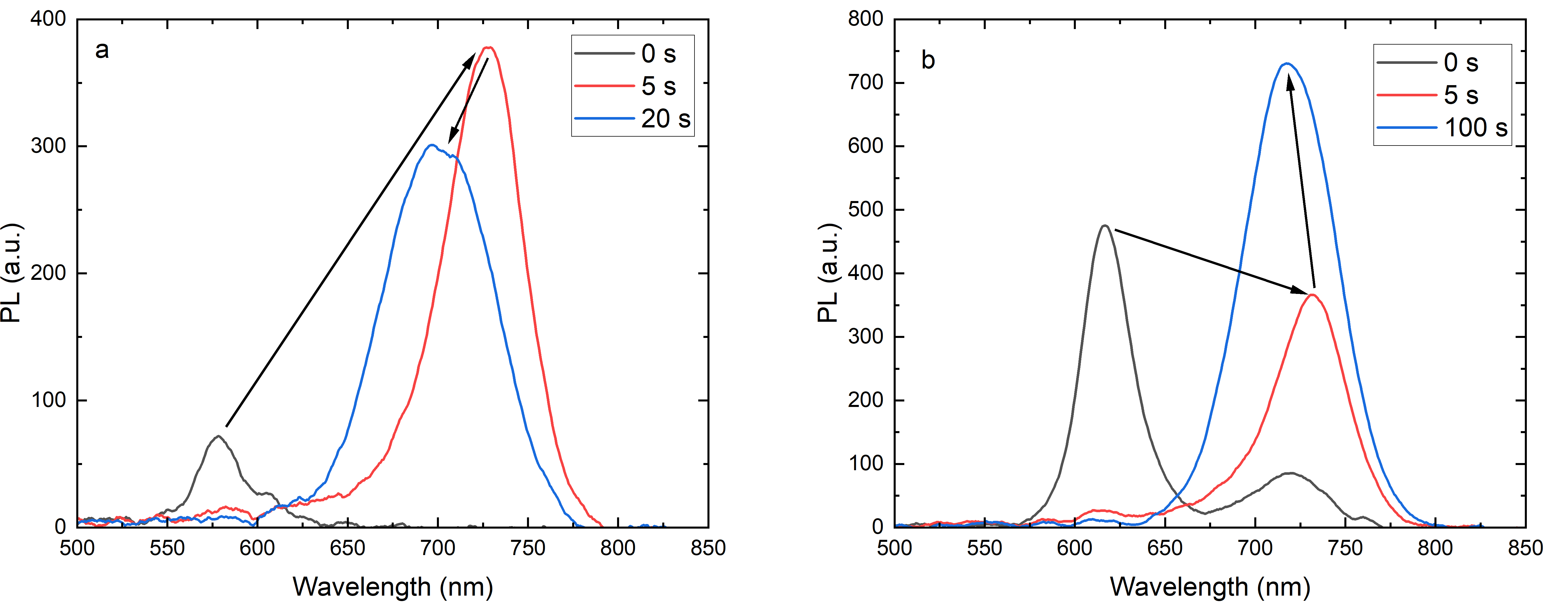}
    \caption{ PL spectra in three time points at room temperature a) MAPbBr$_{2.5}$I$_{0.5}$ b) MAPbBr$_2$I.}
    \label{figs4}
\end{figure}

\begin{figure}
    \centering
    \includegraphics[width=1\textwidth]{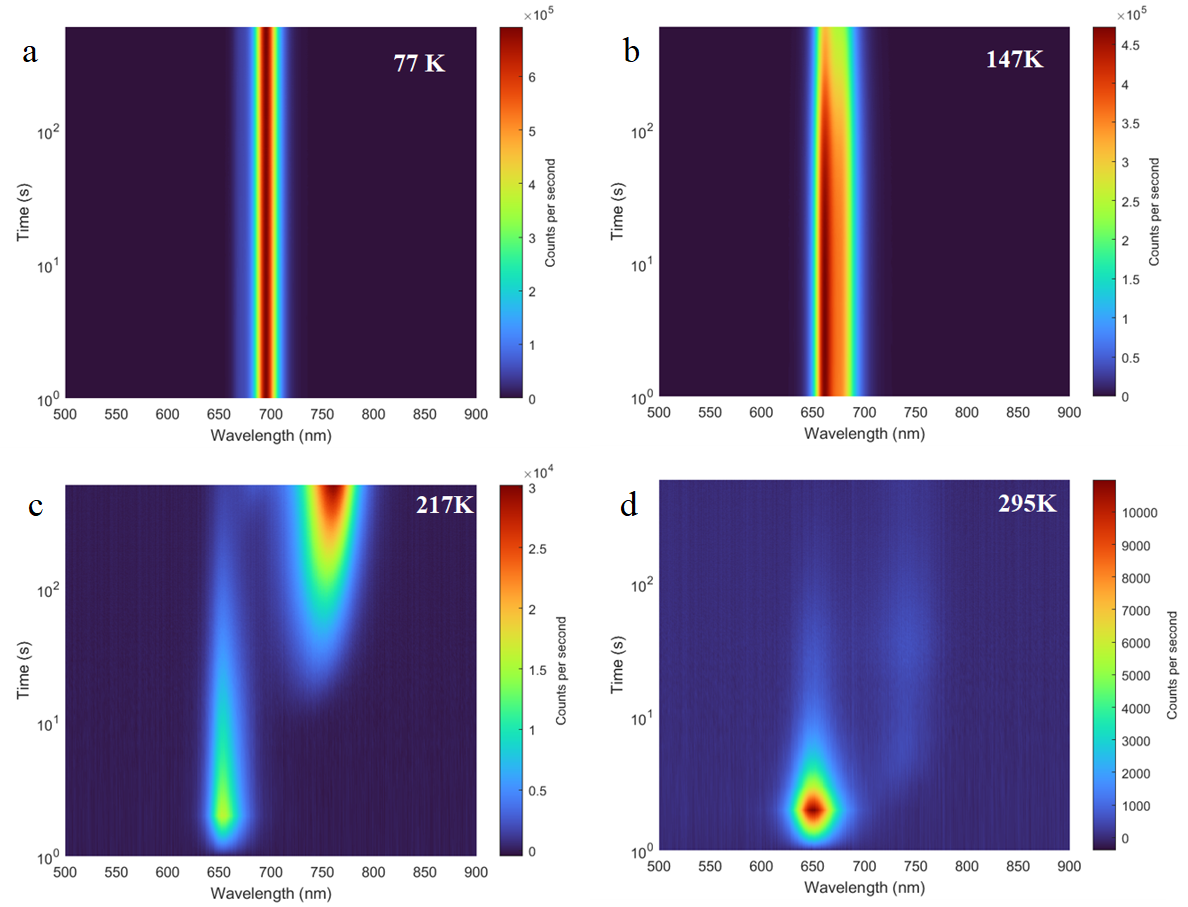}
    \caption{ 2D PL maps for MAPbBr$_{1.5}$I$_{1.5}$ perovskite composition at different temperatures a) 77 k b) 147 K c) 217 K d) 295 K.}
    \label{figs1}
\end{figure}

Lifetimes extrapolation from PL decay was made by bi-exponential fitting (equation 1). Where A and B are the decay amplitudes, D is a constant, x$_0$ is a timing offset, and x represents the decay time constant.

\begin{equation}
    f(x) = A \times exp[-\frac{(x-x_{0})}{t_{rad}}]+B \times exp[-\frac{(x-x_{0})}{t_{srh}}]+D
\end{equation}

\begin{table}
    \centering
    \caption{Radiative and non-radiative lifetimes for MAPbBr$_{2.5}$I$_{0.5}$ perovskite composition at different temperatures.}\label{tab:1}
    \scalebox{1}{
    \begin{tabular}{|l|*{13}{c|}}
        \hline
        \multicolumn{13}{|l|}{\textbf{550 nm - 1000 nm spectral range }} \\ \hline
        T (K) & 77 & 100 & 120 & 140 & 160 & 180 & 200 & 220 & 240 & 260 & 280 & 300 \\
     \hline
        t$_{rad}$ (ns) & 2.14 & 2.21 & 2.28 &	2.07 &	1.9 &	2.07 &	2.39 &	2.18 &	2.27 &	1.96 &	1.92 &	1.55  \\
     \hline
        t$_{srh}$ (ns) & 9.44 & 9.68 & 9.34 & 9.06 & 8.94 & 10.82 & 13.66 &	9.97 &	8.94 &	6.47 & 9.08 & 9.04 \\ 
    \hline
    \multicolumn{13}{|l|}{\textbf{ Initial peak only 550 nm - 650 nm spectral range}} \\ \hline
        T (K) & 77 & 100 & 120 & 140 & 160 & 180 & 200 & 220 & 240 & 260 & 280 & 300 \\
     \hline
        t$_{rad}$ (ns) & - & - & - & - & - & 1.97 &	1.96 &	1.77 &	1.90 &	1.60 &	1.33 &	1.30 \\
        \hline
        t$_{srh}$ (ns) & - & - & - & - & - & 10.75 &	10.54 &	6.85 &	7.36 &	5.83 &	6.60 &	8.95 \\
        \hline
    \multicolumn{13}{|l|}{\textbf{Segregated peak only 650 nm - 1000 nm spectral range}} \\ \hline 
        T (K) & 77 & 100 & 120 & 140 & 160 & 180 & 200 & 220 & 240 & 260 & 280 & 300 \\
     \hline
        t$_{rad}$ (ns) & - & - & - & - & - & 2.69 &	2.74 &	2.19 &	1.91 &	1.96 &	1.90 &	1.55 \\
     \hline
        t$_{srh}$ (ns) & - & - & - & - & - & 13.68 &	13.04 &	9.92 &	7.17 &	6.14 &	9.33 &	9.55 \\
     \hline
    \end{tabular}
    }
\end{table}

\begin{table}
    \centering

    \caption{Radiative and non-radiative lifetimes for MAPbBr$_{2}$I perovskite composition at different temperatures.}\label{tab:2}
    \scalebox{1}{
    \begin{tabular}{|l|*{13}{c|}}
        \hline
        \multicolumn{13}{|l|}{\textbf{550 nm - 1000 nm spectral range }} \\ \hline
        T (K) & 77 & 100 & 120 & 140 & 160 & 180 & 200 & 220 & 240 & 260 & 280 & 300 \\
     \hline
        t$_{rad}$ (ns) & 1.97 &	2.09 &	2.39 &	2.36 &	1.94 &	2.01 &	2.32 &	2.25 &	2.69 &	2.66 &	2.07 &	1.94  \\
     \hline
        t$_{srh}$ (ns) & 5.28 &	5.44 &	5.21 &	5.44 &	4.91 &	5.20 &	6.83 &	7.29 &	9.22 &	9.88 &	7.97 &	5.96 \\ 
    \hline
    \multicolumn{13}{|l|}{\textbf{ Initial peak only 550 nm - 700 nm spectral range}} \\ \hline
        T (K) & 77 & 100 & 120 & 140 & 160 & 180 & 200 & 220 & 240 & 260 & 280 & 300 \\
     \hline
        t$_{rad}$ (ns) & - & - & - & - & - & - &	2.11 &	1.87 &	2.05 &	1.52 &	1.23 &	0.82 \\
        \hline
        t$_{srh}$ (ns) & - & - & - & - & - & - &	6.57 &	5.50 &	6.76 &	6.57 &	5.45 &	3.25 \\
        \hline
    \multicolumn{13}{|l|}{\textbf{Segregated peak only 700 nm - 1000 nm spectral range}} \\ \hline 
        T (K) & 77 & 100 & 120 & 140 & 160 & 180 & 200 & 220 & 240 & 260 & 280 & 300 \\
     \hline
        t$_{rad}$ (ns) & - & - & - & - & - & - &	3.58 &	3.92 &	4.07 &	2.95 &	1.99 &	1.51 \\
     \hline
        t$_{srh}$ (ns) & - & - & - & - & - & - &	7.44 &	11.05 &	12.27 &	10.58 &	7.32 &	5.05 \\
     \hline
    \end{tabular}}
\end{table}

\end{spacing}
\end{document}